# Privacy-Preserving Querying in Sensor Networks[*]

Emiliano De Cristofaro[1], Xuhua Ding[2], and Gene Tsudik[1]

[1] Information and Computer Science, University of California, Irvine, 92697 - {edecrist,gts}@ics.uci.edu
[2] Information Systems, Singapore Management University, 80 Stamford Rd, 178902 - xhding@smu.edu.sg

**Abstract.** Wireless Sensor Networks (WSNs) provide sensing and monitoring services by means of many tiny autonomous devices equipped with wireless radio transceivers. As WSNs are deployed on a large-scale and/or on long-term basis, not only traditional security but also privacy issues must be taken into account. Furthermore, when network operators offer on-demand access to sensor measurements to their clients, query mechanisms should ideally leak neither client interests nor query patterns. In this paper, we present a privacy-preserving WSN query mechanism that uses standard cryptographic techniques. Besides preventing unauthorized entities from accessing sensor readings, it minimizes leakage of (potentially sensitive) information about users' query targets and patterns.

## 1 Introduction

Wireless sensor networks (WSNs) are emerging as a practical and powerful approach to a wide range of monitoring problems. WSN nodes can collectively monitor physical or environmental conditions, such as temperature, sound, vibration, pressure, motion or pollution [32]. However, due to limited resources (computation, storage, bandwidth and, especially, battery power) operations on sensors must be carefully designed. Moreover, once deployed, sensors might operate in a hostile environment and can thus run the risk of various attacks. Besides the usual security issues, WSNs also face the problem of minimizing the security-related overhead, in terms of computation, memory and bandwidth. Over the last decade, much effort has been put into lightweight security techniques geared for WSNs. Promising results have been obtained in key management [16, 26], secure routing (see Section 2), authentication [31], as well as secure data aggregation [9, 8]. For an overview of security issues in WSNs, we refer to [30].

At the same time, as WSNs grow in popularity, the issues of privacy are beginning to surface. So far, it has been often assumed that the information gathered by WSN nodes is ultimately destined for the owner of the network represented by a base station or a sink. However, this assumption does not necessarily hold. The owner and the operator of the network could well be two different entities, similarly to the recent trend of the Database-As-Service model and Software-as-a-Service model (SaaS).

One likely implication is that the WSN is somehow connected to the global Internet via some kind of a gateway and offers data reading/collection service to commercial users.

We consider, for instance, a large long-term WSN deployed for measuring urban pollution levels. This WSN might be owned by the city but its operation can be outsourced to a private contractor. The owner (city) might certify WSN nodes and even initialize them securely. However, day-to-day operation is out of its hands. The contractor provides and manages the interface to the Internet. Moreover, all queries to the WSN must pass through the contractor-owned gateway. A client can request readings from particular sensors at particular locations. As a result, examining privacy threats from the client's perspective becomes both important and interesting. Sensitive information includes nodes or locations queried and the frequency of querying a specific location. We consider as another example the Ocean Tracking Network [3], launched in June 2008, a $168-million project using sensor networks to track movements and behavior of several marine species in 14 ocean regions. Although the project is managed by scientists in Canada, the operators of the network are different global private partners. Preserving privacy becomes a desirable feature of such sensor networks, whenever scientists aim to conduct confidential researches, or users collect readings for business

---

purposes. Within this setting, the network operator becomes a severe threat, since it participates in the data collection process.

A naïve approach to address this problem would be for the client to always query the entire set of sensors. This would achieve perfect privacy but would result in a massive waste of energy. Clearly, efficiency and resource constraints must be taken into account. Another natural solution would be to constantly collect all sensor readings in real time, store them at some (external) server and allow clients to query the database on that server. Privacy issues would then be reduced to the well-known Private Information Retrieval (PIR) problem [12]. However, though popular in the cryptographic research literature, the state-of-the-art in PIR has not reached the point of being practical.

**Contributions.** This paper presents a privacy-preserving query mechanism that allows an external client to collect readings from sensors of his interests without leaking their identities to adversaries including the network operator. Our scheme is based on a technique similar to onion routing [33]. However, it is necessary to overcome several obstacles in order to integrate onion routing with sensor networks, such as the constraints of computation overhead and the packet length, or the presence of multiple malicious nodes in the network. We analyze the security and performance of the mechanism and provide analytical results to show both its effectiveness and its limited overhead.

The rest of the paper is organized as follows. We discuss related work in the next section. Our system setting is described Section 3, followed by the problem definition in Section 4. Section 5 presents a new privacy-preserving query protocol for generic WSNs. Its security and performance are analyzed in Section 6. We then present two extensions in Section 7. The results are summarized in Section 8.

## 2 Related Work

Techniques for anonymous communication across untrusted networks date back to Chaum's pioneering work on MixNets [10]. Today's state-of-the-art is exemplified by TOR [14]. TOR is geared for low-latency synchronous communication over the Internet and is based on the so-called *Onion Routing* principle [33]. In this scheme, each router can only identify the previous and the next hop along a route. However, the sequence of routers must be selected and fixed at connection setup. This task is performed by a special entity (TOR proxy).

Other notable prior work focused specifically on MANETs or WSNs. Secure Dynamic Distributed Routing (SDDR) protocol [15] uses Onion Routing to achieve route anonymity, but does not protect anonymity of source and destination. Anonymous On-Demand Routing (ANODR) is geared for ad hoc networks [22]; it supports untraceable routes or packet flows, with a route pseudonymity approach which decouples a node's location from its identity. ANODR is based on "broadcast with trapdoor information": by embedding certain trapdoor information (known only to the receiver), data is delivered to the receiver but not to other members in the group. However, the identity of the destination is disclosed to nodes along the route and these nodes also gain knowledge of the number of hops from the source. Other interesting results include: SDAR [6] and MASK [39]. However, as pointed out in [35], all these schemes – including ANODR and SDDR – have a common drawback: a passive adversary monitoring all network traffic (without knowledge of any keys) can trace the route reply (RREP) message from destination back to the source.

There are also some prior results in the area of privacy-awareness in location-based services, ubiquitous environments and wireless sensor networks, such as [18, 29], where attempts are made to avoid disclosure of the node's location.

Other related work aims to achieve anonymity in WSNs by obfuscating nodes' identifiers. The work in [37] is the first attempt in this direction, with the assumption that sensors are anonymous, i.e., a sensor has no initial unique identifier. Nodes are organized in clusters and are aware of their own location. The attacker wants to identify and eliminate the minimum number of nodes to inflict maximum loss of data. Two anonymous schemes for clustered WSNs are proposed in [27] to provide anonymity, assuming secret keys shared between sensors and the base station are not compromised. Moreover, nodes in a cluster are considered indistinguishable. Anonymous routing is performed by using pseudonyms to keep node identity private in route requests and replies. In [28], two anonymity techniques are proposed, both based on one-way keyed hash chains. It is assumed that every sensor is aware of its own location and communicates it to the base station. An adversary may eavesdrop on communication and may compromise sensors (to obtain data and individual as well as pairwise encryption keys). The first scheme uses a one-way keyed hash chain to generate a sequence of hash values, each serving as a one-time ID; the second scheme proposes a reverse hashing method.

More recently, [7] presented a scheme that attempts providing query privacy in WSNs. The scheme uses two servers. If these two servers collude, privacy is completely compromised. Furthermore, each sensor needs to store a key for each client, thus resulting in a linear increase in storage overhead. Finally, a scheme for anonymous data collection in WSNs was presented in [20]. The security of this scheme is based on data perturbation[1], rather than on provably secure cryptographic techniques.

In addition, related research was performed in mid-90s to secure *itinerant mobile agents* computation. A mobile agent can be viewed as a process moving from one site to another, preserving its state and data. A mobile agent can autonomously decide when and where to move. Mobile agents can be used to complete a set of user tasks in an asynchronous, dynamic and heterogeneous environments. For instance, a mobile agent can be used to perform an online ticket reservation. The user launches the agent which migrates from site to site in order to find the best price. An overview of this paradigm – along with associated security issues – is given in [11]. However, no practical solutions were proposed.

Another related work in [24] aims to provide both anonymity (of both sender and receiver) and unlinkability. Since mobile agents can self-navigate through the network, the standard onion routing techniques do not apply, since a route is not known at the source. However, in our scenario, the route is chosen at the start by the client who knows the topology of the WSN. Moreover, [24] provides no details about key management.

Finally, directed diffusion [21] has been employed in WSN to globally address a query to all sensors. However, in our setting such solutions would leak information to malicious nodes near the gateway.

## 3 The Setting

We consider a WSN consisting of $n$ nodes that we denote by: $\{S_1, \cdots, S_n\}$. The owner of the WSN is referred to as $\mathcal{OWN}$ and the operator as $\mathcal{OPR}$. As mentioned earlier, these are two distinct entities. Once a sensor obtains environment measurements, it can store the results locally (at least for a short while).

Before deploying the WSN, $\mathcal{OWN}$ chooses a symmetric key encryption scheme $E(\cdot)$, e.g., Rijndael [13]. It also initializes a set of parameters for a light-weight and bandwidth-efficient public key scheme [4] denoted by $\mathcal{EP}(\cdot)$. One very appropriate choice of $\mathcal{EP}(\cdot)$ is the 160-bit Elliptic Curve Integrated Encryption Scheme (ECIES) [4], which we present in Appendix.

Next, $\mathcal{OWN}$ assigns to each $S_i$ a public key $y_i$ and a private key $x_i$ for $\mathcal{EP}(\cdot)$. Then, acting as a Certification Authority (CA), $\mathcal{OWN}$ issues to each $S_i$ a certificate binding $S_i$ to $y_i$.[2] All other tasks are delegated to $\mathcal{OPR}$. At setup time, $\mathcal{OPR}$ publishes the network topology, and node identities: $\{S_1, \cdots, S_n\}$ along with their respective locations and certified public keys. In addition, we assume that each $S_i$ shares a unique symmetric key with its adjacent neighbors.

The network operates in a read-on-demand model, i.e., it responds to external clients' queries. Clients reach the WSN through a gateway, denoted by $\mathcal{GW}$, which belongs to $\mathcal{OPR}$.

To obtain data from a certain sensor, a user (Alice) forms a query and sends it to $\mathcal{GW}$, which, in turn, communicates with the relevant sensor(s). The queried sensor performs requested measurements and returns the results to Alice, again, via $\mathcal{GW}$.

## 4 The Privacy Problem

We decouple security and privacy issues. In our context, the former includes data secrecy, data integrity and authentication. They apply to data provided by the queried sensor: data must be read only by the requesting client; it cannot be replayed, modified or manufactured by the adversary, denoted by $\mathcal{ADV}$. Techniques to provide these security services are standard and do not present a challenge.

---

[1] Sample of the data complement is transmitted to the base station instead of the actual data.
[2] In fact, it is probably just as easy for $\mathcal{OWN}$ to issue a single "communal" certificate to all sensors.

The privacy problem is more difficult. We need to ensure that the identity of the queried sensor remains secret to $\mathcal{ADV}$. Recall that, in our case, $\mathcal{OPR}$ owns $\mathcal{GW}$ and is thus a potential adversary. As a consequence, we also need to keep the content of the reply secret, but for reasons different than simply thwarting unauthorized eavesdroppers. In practice, data collected by sensors is not random; it might depend on the sensor's location. Therefore, if $\mathcal{ADV}$ is allowed to see the reply data, it could determine the location of that data's origin and, hence, the target sensor.

We do not consider privacy of the external client, i.e., we assume that Alice is unconcerned about $\mathcal{ADV}$ knowing that she is querying the WSN. (If this is not the case, techniques like TOR [14] can readily handle building an anonymous channel between Alice and $\mathcal{GW}$.) Finally, we note that denial of service (DoS) attacks are out of scope of this paper.

### 4.1 Portrait of the Adversary

We consider an adversary $\mathcal{ADV}$ who, as mentioned above, controls $\mathcal{OPR}$ and $\mathcal{GW}$. In addition, we allow $\mathcal{ADV}$ to compromise up to $z$ sensors where, $0 \leq z \leq n$. We call these $z$ sensors *infected*. Once a sensor is infected, $\mathcal{ADV}$ learns all of its secret keys, and can thus obtain the plaintext of all incoming and outgoing messages.

Although it controls both $\mathcal{GW}$ and $z$ sensors, we only consider an **honest-but-curious** $\mathcal{ADV}$ [17]. We assume that each infected sensor can only eavesdrop on packets in its physical proximity. In other words, $\mathcal{ADV}$ behaves as a normal sensor during protocol execution (though, it may misuse acquired information). Of course, this generally passive stance does not prevent $\mathcal{ADV}$ from playing the role of some legitimate external user and issuing its own queries.

### 4.2 Our Goal

The goal of our work is to keep the identity of the queried sensor secret. By monitoring many of Alice's queries, $\mathcal{ADV}$ should only gain a negligible advantage in determining the target sensor. Obviously, if $\mathcal{ADV}$ makes a random guess, its success probability should be is $1/n$, assuming that query targets are uniformly distributed across the whole WSN. Note that the number of infected nodes has a significant implication on the security notion and on choosing the appropriate approaches. Two extreme scenarios are as follows:

- $[z = 0]$ – $\mathcal{ADV}$ controls nothing other than $\mathcal{GW}$. In such scenario, $\mathcal{ADV}$ can not eavesdrop on any inter-sensor communication. Since $\mathcal{ADV}$ is at its weakest, solutions can be quite efficient. For instance, one simple approach is to link-encrypt all traffic between $\mathcal{GW}$ and all other sensors on the path. We note that this is exactly the level of privacy offered by [7].
- $[z = n]$ – $\mathcal{ADV}$ controls all sensors as well as $\mathcal{GW}$. Namely, $\mathcal{ADV}$ monitors all traffic in the WSN. This requires the strongest privacy protection. The situation bears a lot of resemblance to PIR [12]. It is thus expected (and likely unavoidable) that schemes offering this level of privacy are impractical, because the computation complexity of any solution is at least linear in terms of WSN size. However, in this worst case, *k-anonymity* can be achieved.[3]

This paper considers a more realistic scenario where $0 \ll z \ll n$. In other words, $\mathcal{ADV}$ controls a fraction of sensors. We argue that, if the WSN coverage area is small, $\mathcal{ADV}$ already has enough *a priori* knowledge of user queries. Moreover, for a small-scale WSN, a trivial privacy approach is to query all nodes, as in a naïve PIR scheme. On the other hand, a large-scale WSN could consist of thousands of sensors, each covering a relatively small area. It is reasonable to assume that $\mathcal{ADV}$ cannot compromise enough sensors to fully cover the entire area covered by the WSN.

---

[3] We recall that k-anonymity is defined in [36] as follows: *... A release provides k-anonymity protection if the information for each person contained in the release cannot be distinguished from at least k-1 individuals whose information also appears in the release.*

### 4.3 Challenges

Our scheme is built upon the onion routing technique whereby sensors play the role of an onion router. Though this technique has been intensively studied for Internet-based communications, we have to carefully tackle the following issues to fully solve the problem above.

- **Data Collection**: The reading from a sensor needs to be returned to a client. In the Internet-based onion routing, the receiver can reply to the sender by constructing a new onion with data enclosed. Such a model is not applicable in sensor networks, because it is too expensive for a sensor to construct an onion. Therefore, the queried sensor has to inject the reading and forward it to the client, without compromising the privacy.
- **Onion Size**: In an Internet onion routing scheme, a node can forward the onion to its next hop via a TCP connection, which does not has any restriction on the onion size. In sensor networks, there is no TCP-like reliable transport protocol. It is expensive for a sensor to buffer packets and application data is always encapsulated into a single packet. Therefore, our protocol does not allow an onion larger than the maximum packet size, which is 128 bytes if using the IEEE 802.15.4 standard. Such a constraint affects the maximum number of layers an onion can have.
- **Energy Consumption**: As in many other sensor network applications, it is among the top priorities to preserve the sensor's energy. This restricts the sensor's cost for both the computations and the communications, i.e. receiving and transmitting an onion.

## 5 Basic Construction

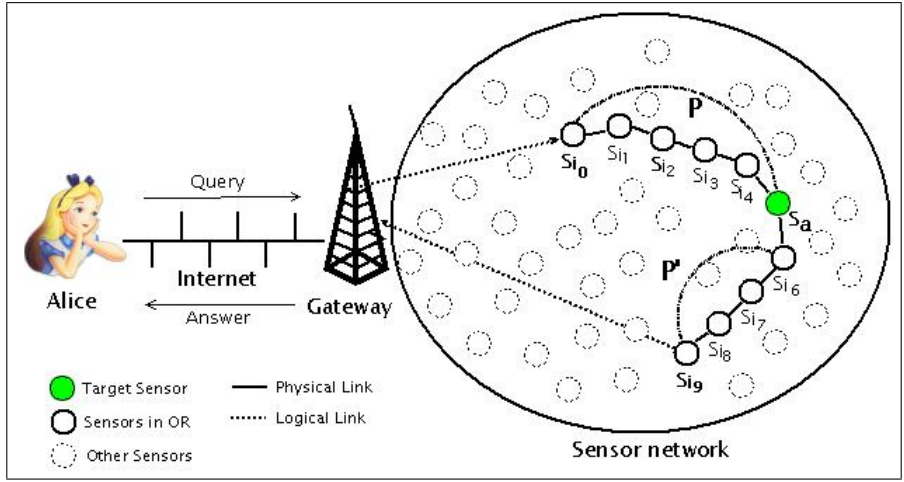

**Fig. 1.** Route Example with $t = 8$

The basic idea behind our scheme is to provide anonymity using an onion routing-like technique. To obtain better privacy, we require for each node in the onion route to contribute its own reading. In current onion routing solutions (e.g., TOR), the last hop on the onion route is the destination. This information leakage should be avoided in our protocol due to the presence of *infected* nodes.

We assume that Alice knows the topology of the network. However, we argue that such assumption is realistic in the applications we consider (cf Section 1). In scenarios such as the urban pollution measurement service, the user is fully aware of the nodes in the network although the network itself is operated by an external contractor.

Alice selects a route which includes the target and a set of other randomly chosen sensors. Both the first and the last hop of the route are set to $\mathcal{GW}$. Alice constructs the query by performing a number of encryptions, each analogous to an onion layer. The query is then forwarded to $\mathcal{GW}$. In the end, Alice receives a reply (through $\mathcal{GW}$) which she processes to obtain the desired measurements.

The WSN query mechanism is composed of three algorithms: 1) **query generation**, 2) **data collection**, and 3) **data recovery**. Alice generates a query and sends it to $\mathcal{GW}$, which runs the data collection algorithm and returns the outcome to Alice. Alice then runs the data recovery algorithm to obtain the readings.

In the rest of the paper, we use the notation in the following Table.

| Symbol | Meaning |
|---:|---|
| $n$ | total number of sensors |
| $\mathcal{G}$ | WSN connectivity/topology graph |
| $z$ | number of infected (compromised) sensors |
| $t$ | onion route length in hops; (usually related to the WSN diameter) |
| $m, c$ | plaintext and ciphertext |
| $S_i$ | $i$-th sensor; $0 \leq i < n$ |
| $y_i, x_i$ | $S_i$'s public and private keys |
| $\mathcal{EP}_j(m)$ | public key encryption ( of message m with key $y_j$ ) |
| $\mathcal{DP}_j(c)$ | public key decryption ( of ciphertext c with key $x_j$ ) |
| $\mathcal{E}(k, m)$ | symmetric key encryption ( of message m with key $k$ ) |
| $\mathcal{D}(k, c)$ | symmetric key decryption ( of ciphertext c with key $k$ ) |
| $L_r$ | bit size of one sensor reading |
| $L_c$ | bit size of onion header |

### 5.1 Query Generation

Suppose that Alice wants to query a sensor $S_a$. The query consists of two parts:

1. *Header:* an onion-like structure which governs how sensors forward and respond to the query. The header has $t + 1$ layers, each corresponding to a sensor hop on the route, where $t$ is a *fixed* public parameter chosen by Alice. (We expect $t$ to be closely related to WSN diameter). The size of the header is also fixed – $L_c$ bits – as the maximum WSN packet length.
2. *Body:* a random-looking binary string that carries the readings back to Alice. It is comprised of $t$ slots, each containing an $L_r$-bit reading. The positions of sensor readings are randomly permuted by Alice. (As described later, each onion layer tells the corresponding sensor the slot index within the *body* where to deposit its encrypted reading.) This way, an intermediate node is unaware of its position in the route.

**Path Selection.** Using the publicly known WSN topology, Alice selects a random path (containing $S_a$) of length $t + 2$:
$$\langle S_{i_0}, S_{i_1}, \cdots, S_{i_t}, S_{i_{t+1}} \rangle$$
where all pairs $S_{i_j}, S_{i_{j+1}}$ ($0 \leq j \leq t+1$) are immediate neighbors and, for some $v \in [1, t]$, $S_{i_v} = S_a$. However, $\mathcal{GW}$ and $S_{i_0}$ (as well as $S_{i_{t+1}}$ and $\mathcal{GW}$) are not required to be adjacent[4]. In other words, Alice constructs a $(t+3)$-hop circuit which starts with $\mathcal{GW}$, runs through $(t+2)$ sensors (among which is $S_a$) and ends back at $\mathcal{GW}$. See Figure 1 for an example. The path selection process is shown in more detail in Algorithm 1.

**Header Construction.** First, Alice creates a random permutation of indices $\{1, 2, \cdots, t\}$ which we denote as: $\pi = \{p_1, p_2, \cdots, p_t\}$. Each $p_j$ is used to inform $S_{i_j}$ ($j$-th node in the path) where to insert its reading within the body. Furthermore, we define $p_0 = 0$ and $p_{t+1} = 0$. This is to instruct the end-points of the path ($S_{i_0}$ and $S_{i_{t+1}}$) that their readings are not needed, since they are both visible to $\mathcal{GW}$. Next, Alice generates a set of $t$ random symmetric (e.g., Rijndael) keys: $k_1, ..., k_t$ and sets $k_0 = k_{t+1} = 0$.

---
[4] As the end-points of the path, $S_{i_0}$ and $S_{i_{t+1}}$ are assumed to be "visible" (known) to $\mathcal{GW}$.

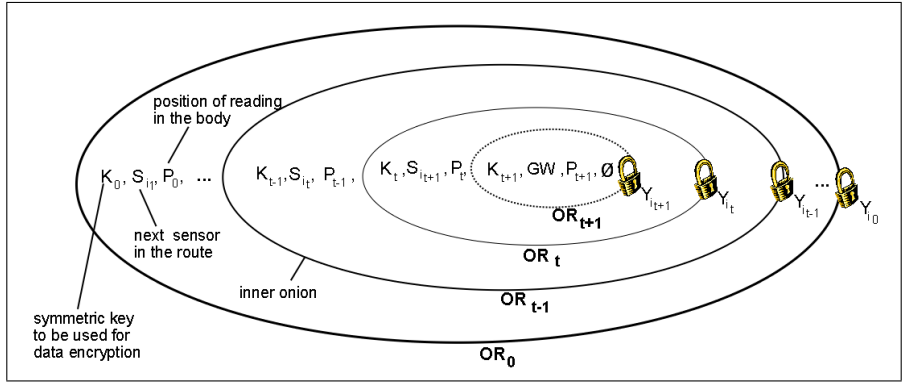

**Fig. 2.** The final onion (header) built by Alice.

She then constructs the header by setting $OR_{t+2} = \emptyset$ and encrypting $< k_{t+1}, \mathcal{GW}, p_{t+1}, OR_{t+2} >$ with $y_{i_{t+1}}$, i.e. $S_{i_{t+1}}$'s public key. Let:

$$OR_{t+1} = \mathcal{EP}_{i_{t+1}}(\langle k_{t+1}, \mathcal{GW}, p_{t+1}, OR_{t+2} \rangle )$$

denote the resulting ciphertext.
Then, she encrypts $< k_t, S_{i_{t+1}}, p_t, OR_{t+1} >$ with $y_{i_t}$, which results in:

$$OR_t = \mathcal{EP}_{i_t}(\langle k_t, S_{i_{t+1}}, p_t, OR_{t+1} \rangle ) \tag{1}$$

Alice repeats this process, until computing $OR_0$ – encryption of $< k_0, S_{i_1}, p_0, OR_1 >$ under $y_{i_0}$.

The resulting final onion is showed in Figure 2.

Since the size of the header grows after each encryption, we apply the technique in [19] to keep ciphertext size constant. This approach will prevent any sensor on the onion route from determining its position within the route. The resulting header is an $L_c$-bit string.

**Body Construction.** The initial body $M_0$ is just a random binary string of size $L_r * t$. During the data collection phase, random bits are gradually replaced with encrypted readings.

The query construction algorithm outputs the initial query $\langle OR_0, M_0 \rangle$. Alice then sends it to $\mathcal{GW}$ which, forwards it intact to $S_{i_0}$. Upon receiving the query, each sensor runs Algorithm 2 and forwards the query to the next hop.

### 5.2 Data Collection

When $S_{i_j}$ receives $\langle OR_j, M_j \rangle$, it decrypts $OR_j$ and obtains: permuted position $p_j$ (in the body) for depositing its reading, next hop $S_{i_{j+1}}$, one-time secret key $k_j$ and inner onion $OR_{j+1}$ destined for $S_{i_{j+1}}$. (Recall that, if $p_j = 0$, no data is solicited.) Details are shown in Algorithm 2. Each sensor in the route performs the same sequence of steps. It is easy to see that the client's actual target sensor $S_a$ can not determine its "special" role.

With every collection step (hop), a layer of the onion is consumed and the size of the *header* decreases after each decryption. The difference in size can be exploited by $\mathcal{ADV}$(or the current-hop sensor), to determine the position of the current hop in the route. To overcome the problem, we use the approach introduced in [19]. Alice introduces a random $\delta$-bit string as padding in the innermost onion and then trims it after every encryption such that the final onion size is always $L_c$-bits. $S_{i_j}$ consumes one layer of the onion and restores its size by appending random bits. This way, neither an eavesdropper nor a sensor in path can derive the

**Algorithm 1**: Path Selection: executed by a WSN client (Alice)

Input: $(S_a, t, \{S_1, \cdots, S_n\}, \mathcal{G})$
1. Generate random $w \in_r \{1, ..., t\}$
2. Let $i_w = a$    /* $S_a$ sits randomly in the path */
3. **for** $(j \leftarrow w - 1$ **to** $0)$ **do**
   Select $S_{i_j}$ as a random neighbor of $S_{i_{j+1}}$
4. Let $P = \{S_{i_0}, S_{i_1}, ..., S_{i_w}\}$ where $S_{i_w} = S_a$
5. If ($P$ contains a loop) go to 1
6. **for** $(j \leftarrow w + 1$ **to** $t)$ **do**
   Select $S_{i_j}$ as a random neighbor of $S_{i_{j-1}}$
7. Let $P' = \{S_{i_{w+1}}, S_{i_{w+2}}, ..., S_{i_t}\}$
8. If ($P'$ contains a loop) OR ($P \cap P' \neq \emptyset$) go to 7
9. Return Path $=<\mathcal{GW}, P, P', \mathcal{GW}>$

---

**Algorithm 2**: Data Collection: executed by $S_{i_j}, 0 \leq j \leq t+1$

Input: $(OR_j, M_j)$
1. Decrypt $OR_j$ using $x_{i_j}$. Abort of failure.
2. Parse decrypted string as: $<k_j, S_{i_{j+1}}, p_j, OR_{j+1}>$.
3. Pad $OR_{j+1}$ with randomness up to $L_c$ bits.
4. **if** $(p_j \neq 0)$ **then**
    Obtain local reading $\alpha_j$
    /*Let $M_j[p_j]$ be the $p_j$-th $L_r$-bit block of $M_j$ */
    Set $M_j[p_j] = \alpha_j$
5. **for** $(s \leftarrow 1$ **to** $t)$ **do**
    /* Let $M_j[s]$ be the $s$-th $L_r$-bit block of $M_j$ */
    Set $M_{j+1}[s] = \mathcal{E}(k_j, M_j[s])$
    /* each block is encrypted separately */
6. Send $\langle OR_{j+1}, M_{j+1}\rangle$ to $S_{i_{j+1}}$

---

current (relative) position in the route from the current header size. (However, the last hop is aware of its position in the route since the next hop is $\mathcal{GW}$.) Finally, upon receipt of $\langle OR_{t+2}, M_{t+2} \rangle$, $\mathcal{GW}$ forwards it intact to Alice. We refer to [19] for the discussion on trimming and padding that keep the onion size constant.

### 5.3 Data Recovery

Alice recovers the reading by unraveling layer-by-layer $M_{t+2}$ received from $\mathcal{GW}$. Recall that $M_{t+2}$ is a concatenation of concentrically-encrypted $t + 2$ blocks of ciphertext. Among them, Alice is only interested in one block at position $p_v$ – the one containing the reading from the target sensor $S_a = S_{i_v}$. According to the data collection algorithm, this block is encrypted by $S_{i_v}$ and super-encrypted by all subsequent hops in the route. There are thus $(t + 2 - v)$ layers of encryption (with keys $\{k_v, ..., k_{t+1}\}$, respectively) which must be stripped off one-by-one in order for Alice to obtain $\alpha_v$. The recovery phase is shown in Algorithm 3.

## 6 Analysis

### 6.1 Security Analysis

We note that observing the onion does not reveal any information apart from the next hop of the route. After decrypting an incoming onion, a sensor obtains an inner onion. Every onion is in fact the ciphertext of a public key envelope encryption. Therefore, given a semantically secure public key encryption scheme, the onion reveals no information except its length, provided that the sensor does not possess the corresponding

**Algorithm 3**: Data Recovery: executed by a WSN client (Alice).

Input: $(OR_{t+2}, M_{t+2})$
/* Recall that $OR_{t+2} = \emptyset$ */
1 Let $TMP = M_{t+2}[p_v]$
/* $p_v$-th $L_r$-bit block of $M_{t+2}$ */
2 **for** $(s \leftarrow t+1$ **to** $v)$ **do**
  Set $TMP = \mathcal{D}(k_s, TMP)$
  /* $k_v$ is the last decryption key */
3 Parse $TMP$ as: $<\alpha_v>$
/* abort if error */
4 Output $\alpha_v$
/* reading from target sensor $S_{i_v} = S_a$*/

---

private key. Moreover, all sensors in the route perform the same operation. Thus, even knowledge of a sensor's position in the onion route does not yield the location of the actual target sensor.

**Number of Exposed/Known Nodes.** We now analyze the information that $\mathcal{ADV}$ obtains by observing a node in the onion route.

$\mathcal{ADV}$ cannot recover any information about the previous node since no sender information is present in most WSN routing protocols. However, $\mathcal{ADV}$ can easily see the next hop since destination is visible. This way, for every infected node, $\mathcal{ADV}$ learns the identity of two hops in the onion route.

We introduce the term *known* nodes, to describe the set of hops in the route known to $\mathcal{ADV}$. Note that not all of them are directly infected. From now on, we denote with $x$ the number of known nodes, and we show that it can be expressed it in terms of $n$, $z$ and $t$.

To compute the number of expected known nodes, we consider, for every node in the onion: (i) the probability that the node is infected, i.e., $\frac{z}{n}$, (ii) the probability that the node is not infected but its predecessor in the onion is, i.e., $\frac{z}{n} \cdot \frac{n-z}{n}$.

As a result, the expected number of *known* nodes among $t$ onion sensors is:

$$\mathrm{E}(x) = t \cdot \left( \frac{z}{n} + \frac{z}{n} \cdot \frac{n-z}{n} \right) \tag{2}$$

We remark that the number in (2) has also been validated by running the following experiment. We simulated sensor networks of different sizes $n$, where we selected *z-out-of-n* nodes to simulate infected nodes. Finally, we counted the number of known nodes. (We recall that a node is "known" if the node itself or its predecessor in the route is infected). The experiments were repeated 10,000 times to average the values, which proved to match our analytical formula in (2). (Results are not enclosed due to lack of space).

**Target Sensor Privacy.** Our goal is to protect the privacy of the target sensor. To this aim, we compute the *probability of $\mathcal{ADV}$ breaking the query privacy: $\mathcal{ADV}$ successfully guesses which sensor in the network is the query target*.

We remark that requesting $t$ nodes to report their readings makes the scheme at worst $t$-anonymous. Therefore, the above probability is upper-bounded by $1/t$. However, we argue that our idea to combine onion routing techniques with $t$-anonymity provides much stronger privacy.

The probability of breaking the privacy depends on the different network and adversarial parameters: $z$, the number of infected nodes; $x$, the number of known nodes on the onion route; $n$, the size of the network; $t$, the onion size. When guessing, $\mathcal{ADV}$ can exclude the $z$ infected nodes which are not touched by the route, and use the information of the known nodes on the route.

Hence, the probability of $\mathcal{ADV}$ successfully guessing the query target based on $x$ known nodes is: $\frac{1}{n-z+x-\frac{z}{n} \cdot \frac{n-z}{n} \cdot t}$, hence:

$$\Pr(\mathcal{ADV} \text{ breaks privacy}) = \frac{1}{n - z + t \cdot z/n} \tag{3}$$

To show the effectiveness and the scalability of our privacy solution, we plot the above probability with different adversarial and network settings. Figure 3 shows that for different network sizes, namely $n = 100, 1000$, and $10000$, using a constant size onion route of $t = 20$ hops, the probability of $\mathcal{ADV}$ breaking the privacy scales according to the ratio of infected nodes. The worst case, i.e. $t$-anonimity is reached for 100% of infected nodes.

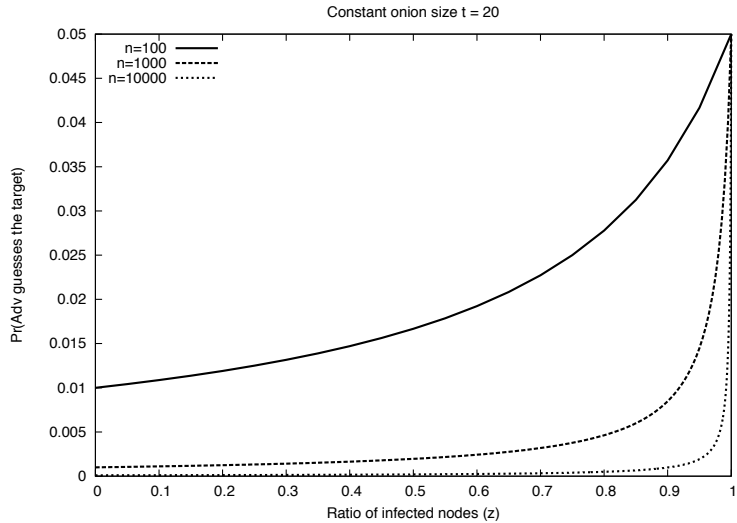

**Fig. 3.** Probability of $\mathcal{ADV}$ breaking query privacy

Finally, we stress that, since the worst case for our scheme is $t$-anonymity, we can let the client select the same route for the same queried node. Thus, $\mathcal{ADV}$ can not intersect its views of query executions.

### 6.2 Performance Analysis

Both query generation and data recovery algorithms are executed by a client (external to the WSN) which has sufficient computing resources to handle standard cryptographic operations. Therefore, our performance analysis only focuses on sensor energy consumption due the execution of the data collection algorithm and in forwarding an onion.

Using experiment results from the relevant literature, we estimate the computation energy cost of the sensors on the onion route. Note that sensors not on the route do not perform any computation and the expensive onion generation is done by the client instead of a sensor. A sensor on the route performs only one ECC decryption and one Rijndael encryption. We refer to a configurable ECC engine – TinyECC [25]. [2] shows that an Imote2 (13MHz) consumes 13.73 $mJ$ in performing a 160-ECC decryption. An earlier result in [23] measured the cost of symmetric key encryption in sensors. According to it, speed-optimized Rijndael in CBC mode requires around 350 CPU cycles to encrypt a 90-byte message. Based on the Imote2 data-sheet [2], the CPU of Imote2 operates at low voltage (0.85v) and low frequency (13MHz). Thus, we estimate that one Rijndael-CBC decryption of an onion costs roughly 1 $mJ$ (CPU only), which is negligible as compared with the cost of public key (e.g., ECIEC) operations.

Another factor in energy consumption is due to onion forwarding. The maximum size of a TinyOS packet is 128 bytes [1], including 11-byte header and 2-byte CRC. According to the CC2420 data-sheet, for the radio transceiver of Imote2, it costs about $123mK$ and $114mJ$ to receive and transmit a 128-byte packet, respectively. Therefore, the total estimated energy cost for a sensor on the route is roughly $250mJ$.

The proposed scheme significantly saves energy compared to reading all the sensors. Indeed, energy consumption of sensor nodes is affected mainly by communication rather than computation, and reading all sensors entails significant communication overhead. Compared to the PIR approach mentioned in Section 1, our scheme is also more cost-effective. In our solution, sensors are awakened only when reached by a query. In contrast, the PIR approach awakens all sensors periodically, even if some readings are not requested.

**Onion route length.** The size of the packet limits the query header size, and consequently the onion route length. We use two optimization tricks to minimize the size increase of the query header. The ECIES ciphertext consists of the output from a symmetric key encryption and a 160-bit random number $R$. One optimization is that $R$ can be reused for every ECIES encryption in the query[5], which is provably secure as shown in [5]. The other optimization is that $k_t$ in Equation 1 can be the same symmetric key used in the ECIES encryption for $OR_t$. $S_{i_t}$ can derive it when it deciphers the onion. Note that the ciphertext of a symmetric key encryption has almost the same length as its plaintext. Thus, $OR_j$ is only slightly longer than $OR_{j+1}$ due to the insertion of $S_{i_{j+1}}$ and $p_j$.

The query generation algorithm outputs $(t \cdot L_r + L_c)$-bit onions. Let the sensor identities be $b$ bits long. Therefore, $L_c = 160 + (b + \log t)t$. The total onion route size is thus $160 + (L_r + b)t + t \log t$. Since the payload of a packet is 115 bytes, we have $t = 22$ in a conservative estimation, when $L_r = b = 16$. An Imote2 sensor's antenna covers about 30 meters. For larger onion length, we propose two extended constructions in the next section.

# 7 Extensions

The scheme presented in Section 5 illustrates the basic idea of using onion routing in a sensor network. However, this approach is not always fully scalable to large WSNs since we anticipate the onion route length to be close to the network diameter. Moreover, as the WSN gets larger, the maximum packet size typically remains the same. This limits the number of hops in the onion route. For this reason, we propose two extensions: *overlay onion routing* and *hybrid onion routing*.

## 7.1 Overlay Onion Routing

In this approach, Alice randomly picks a set of $t + 1$ sensors. However, unlike before, these sensors ($S_{i_j}$ and $S_{i_{j+1}}$) are not required to be adjacent. This set of sensors, together with Alice's target sensor $S_a$ (placed at a random position between 1 and $t$) form a $t + 3$-hop overlay onion route starting and ending in $\mathcal{GW}$. The data collection algorithm and the data recovery algorithm remain unchanged. Transmission between two elements of the onion route is obtained via WSN's own routing mechanism. With this approach, the choice of $t$ is independent of the size of WSN.

Nonetheless, as a result, we trade some privacy for efficiency and scalability. In fact, a compromised node residing on the route segment between $S_{i_j}$ and $S_{i_{j+1}}$ can learn $S_{i_{j+1}}$ by simple eavesdropping. Thus, $\mathcal{ADV}$ has a higher probability in learning a hop in an onion route than in the basic scheme. The exact probability advantage depends on the underlying WSN routing protocol. Since messages in a typical WSN are transmitted in a broadcast fashion, this probability is non-negligible. (However, simple link encryption can be used to reduce – but not prevent – such attacks.)

## 7.2 Hybrid Onion Routing

Both the basic scheme and the overlay extension are geared for WSNs composed of homogeneous sensor nodes. The concept of heterogeneous sensor networks [38, ?] has been explored recently in order to improve reliability of data transmissions. This hybrid type of WSN involves a set of powerful *backhaul* nodes which form an overlay network, thus allowing high-bandwidth and highly-reliable communications. We can extend our basic scheme by exploiting the features of hybrid WSNs. The main idea is to use overlay onion routing to forward along a route of randomly chosen backhaul nodes, while these backhaul nodes use our basic scheme to collect data from regular sensors. We omit any further description of this extension since it is rather intuitive.

---
[5] The symmetric key cipher we use is also probabilistic. However, its randomness should not be reused.

### 7.3 Query Pattern Privacy

The basic scheme entails Alice using the same route for repeated queries of a given target sensor. Although this prevents $\mathcal{ADV}$ from learning the location of Alice's target, it exposes Alice's query patterns. In other words, given $m$ queries executed by Alice, $\mathcal{ADV}$ can derive the frequency distribution of Alice's targets, even though their identities are unknown.

One remedy is for Alice to construct a small pool of randomly selected sensors, denoted by $\Gamma$. The set size depends on the size of her target set. When constructing an onion route, Alice selects a fixed number of onion sensors (say, $c$) from $\Gamma$ and $t - c - 1$ sensors from the entire population of sensors. This way, $\mathcal{ADV}$ can not determine whether a duplicate occurrence of a sensor is due to the retrieval from $\Gamma$ or Alice's query pattern.

## 8 Conclusions

This paper proposed a technique for ensuring query privacy for on-demand WSN access. We compared our proposal with related work showing that, to the best of our knowledge, it represents the first mechanism for query privacy in terms of identity, location, and frequency of queried nodes. We introduced a security model, whereby the adversary compromises a fraction of nodes, using them to gain information about query targets and patterns. As part of future work, we plan to implement and test our mechanism in the TinyOS setting and conduct experiments to obtain better assessment of its overhead.

## References

1. CC2420 radio stack. http://www.tinyos.net/tinyos-2.x/doc/txt/tep126.txt.
2. Imote2: High-performance wireless sensor network node. http://www.xbow.com/Products/Product_pdf_files/Wireless_pdf/Imote2_Datasheet.pdf.
3. Ocean Tracking Network. http://www.oceantrackingnetwork.org/.
4. Certicom Research: Standards for efficient cryptography - SEC1: Elliptic curve cryptography. http://www.secg.org/download/aid-385-sec1_final.pdf, 2000.
5. M. Bellare, A. Boldyreva, and J. Staddon. Randomness reuse in multi-recipient encryption schemes. In *Proceedings of PKC'03*, 2003.
6. A. Boukerche, K. El-Khatib, L. Xu, and L. Korba. A novel solution for achieving anonymity in wireless ad hoc networks. In *Proceedings of ACM PE-WASUN'04*, pages 30–38, 2004.
7. B. Carbunar, Y. Yu, L. Shi, M. Pearce, and V. Vasudevan. Query privacy in wireless sensor networks. In *Proceedings of SECON'07*, 2007.
8. C. Castelluccia, E. Mykletun, and G. Tsudik. Efficient aggregation of encrypted data in wireless sensor networks. In *Proceedings of Mobiquitous'05*, pages 109–117, 2005.
9. H. Chan, A. Perrig, and D. Song. Secure hierarchical in-network aggregation in sensor networks. In *Proceedings of ACM CCS'06*, pages 278–287, 2006.
10. D. L. Chaum. Untraceable electronic mail, return addresses, and digital pseudonyms. *Communications of ACM*, 24(2):84–90, 1981.
11. D. Chess, B. Grosof, C. Harrison, D. Levine, C. Parris, and G. Tsudik. Itinerant agents for mobile computing. *IEEE Personal Communications*, pages 267–282, 1998.
12. B. Chor, E. Kushilevitz, O. Goldreich, and M. Sudan. Private information retrieval. *Journal of ACM*, 45(6):965–981, 1998.
13. J. Daeman and V. Rijmen. AES proposal: Rijndael. 1999.
14. R. Dingledine, N. Mathewson, and P. Syverson. TOR: The second-generation onion router. In *Proceedings of 13th USENIX Security Symposium*, volume 2, 2004.

15. K. El-Khatib, L. Korba, R. Song, and G. Yee. Secure dynamic distributed routing algorithm for ad hoc wireless networks. In *Proceedings of ICPPW'03*, pages 359–366, 2003.
16. L. Eschenauer and V. D. Gligor. A key-management scheme for distributed sensor networks. In *Proceedings of ACM CCS'02*, pages 41–47, 2002.
17. O. Goldreich. *Foundations of Cryptography: Basic Tools*. Cambridge University Press, New York, NY, USA, 2000.
18. M. Gruteser, G. Schelle, A. Jain, R. Han, and D. Grunwald. Privacy-aware location sensor networks. In *Proceedings of HOTOS'03*, pages 28–32, 2003.
19. C. Gulcu and G. Tsudik. Mixing email with babel. In *Proceedings of NDSS'96*, pages 2–16, Washington, DC, USA, February, 1996. IEEE Computer Society.
20. J. Horey, M. M.Groat, S. Forrest, and F. Esponda. Anonymous Data Collection in Sensor Networks. In *Proceedings of Mobiquitous'07*, 2007.
21. C. Intanagonwiwat, R. Govindan, and D. Estrin. Directed diffusion: a scalable and robust communication paradigm for sensor networks. In *Proceedings of MobiCom'00*, pages 56–67, 2000.
22. J. Kong and X. Hong. ANODR: Anonymous on demand routing with untraceable routes for mobile ad-hoc networks. In *Proceedings of MobiHoc'03*, pages 291–302, June, 2003.
23. Y. Law, J. Doumen, and P. Hartel. Survey and benchmark of block ciphers for wireless sensor networks. *ACM Transactions on Sensor Networks*, 2(1), 2006.
24. R. Leszczyna and J. Górski. Untraceability of mobile agents. In *Proceedings of AAMAS'05*, pages 1233–1234, July, 2005.
25. A. Liu and P. Ning. TinyECC: A configurable library for elliptic curve cryptography in wireless sensor networks. In *Proceedings of the ICIPSN'08*, pages 245–256, 2008.
26. D. Liu, P. Ning, and R. Li. Establishing pairwise keys in distributed sensor networks. *ACM Transaction on Information System Security*, 8(1):41–77, 2005.
27. S. Misra and G. Xue. Efficient anonymity schemes for clustered WSN. *International Journal of Sensor Networks*, 1(1):50–63, 2006.
28. Y. Ouyang, Z. Le, Y. Xu, N. Triandopoulos, S. Zhang, J. Ford, and F. Makedon. Providing Anonymity in Wireless Sensor Networks. In *Proceedings of ICPS'07*, 2007.
29. C. Ozturk, Y. Zhang, and W. Trappe. Source-location privacy in energy-constrained sensor network routing. In *Proceedings of SASN'04*, pages 88–93, 2004.
30. A. Perrig. Security in sensor networks: industry trends, present and future research directions. In *Proceedings of ACM WiSe'06*, pages 53–53, 2006.
31. A. Perrig, R. Szewczyk, V. Wen, D. Culler, and J. D. Tygar. SPINS: Security protocols for sensor networks. *Wireless Networks*, 8(5):521–534, 2002.
32. C. S. Raghavendra, K. M. Sivalingam, and T. Znati. *Wireless Sensor Networks*. 2004.
33. M. G. Reed, P. F. Syverson, and D. M. Goldschlag. Anonymous connections and onion routing. *IEEE Journal on Selected Areas in Communications, Special Issue on Copyright and Privacy Protection*, 16(4):482–494, 1998.
34. T. Sander and C. F. Tschudin. Protecting mobile agents against malicious hosts. *Mobile Agents and Security*, pages 44–60, 1998.
35. S. Seys and B. Preneel. Arm: Anonymous routing protocol for mobile ad hoc networks. In *Proceedings of AINA'06*, pages 133–137, 2006.
36. L. Sweeney. k-Anonymity: A model for Protecting Privacy. *Int. J. Uncertain. Fuzziness Knowl.-Based Syst.*, 10(5):557–570, 2002.
37. A. Wadaa, S. Olariu, L. Wilson, M. Eltoweissy, and K. Jones. On providing anonymity in wireless sensor networks. In *Proceedings of ICPADS'04*, pages 411–418, 2004.
38. M. Yarvis, A. Kushalnagar, H. Singh, Y. Liu, and S. Singh. Exploiting heterogeneity in sensor networks. In *Proceedings of Infocom*, 2005.
39. Y. Zhang, W. Liu, and W. Lou. Anonymous communications in mobile ad hoc networks. In *Proceedings of IEEE INFOCOM'05*, 2005.

## Appendix A: ECIES

ECIES [4] is a well-known public key encryption scheme geared for resource-constrained devices, such as sensors. According to [4], 160-bit ECIES provides roughly the same level of security as 1024-bit RSA encryption. An ECIES key-pair is usually denoted as $(x, Y)$, where $x$ is the secret key and $Y = xP$ is the public key. $P$ is a base point on the elliptic curve defined over $\mathbb{F}_q$ where $q$ a 160-bit prime.

To encrypt a message $m$ using $Y$, the sender first generates an *ephemeral* key-pair $(r, R)$ where $r \in_R \mathbb{Z}_q^*$ and $R = rP$. Then, it computes $rY$ and obtains a symmetric encryption key $k$ using a key derivation function (KDF). It then computes $C = E_k(m)$ where $E_k()$ denotes symmetric encryption under $k$. The final ECIES ciphertext consists of $R$, $C$ and certain integrity checking information. To decrypt, the receiver (who knows the private key $x$) computes $xR$, which allows it to derive $k$ and thus recover $m$ from $C$.